\documentclass[fleqn,twoside,a4paper]{article}
\usepackage{amsmath}
\usepackage{espcrc2}
\usepackage{bm}
\usepackage[dvips]{color,graphicx}
\newcommand{\ud}     {\mathrm{d}}
\newcommand{\gev}    {\;\mathrm{GeV}}

\newcommand{\gevsq}  {\;\mathrm{GeV}^2}

\newcommand{\ceps}{\varepsilon}

\newcommand{\eq}[1]{Eq.(\ref{#1})}

\title{Phenomenology of lepton-nucleus DIS} 
\author{S. A. Kulagin%
\address{
Institute for Nuclear Research, 117312 Moscow, Russia}%
\thanks{Supported in part by the Russian Foundation for 
Basic Research project 03-02-17177 and INTAS project 03-51-4007.}
and R. Petti%
\address{CERN, CH-1211 Gen\'eva 23, Switzerland}
\address{University of South Carolina, Columbia SC 29208, USA}
}
\begin{document}

\begin{abstract}
\noindent
The results of recent phenomenological studies of unpolarized nuclear
deep-inelastic scattering are discussed and applied to calculate
neutrino charged-current structure functions and cross sections for a
number of nuclei.
\end{abstract}

\maketitle

\section{Introduction and motivations}
\label{sec:intro}

Significant nuclear effects were discovered in charged-lepton (CL)
deep-inelastic scattering (DIS) experiments (for a review see
\cite{Arneodo:1994wf,Piller:1999wx}).
The experimental observations indicate that the nuclear environment plays 
an important role even at energies and momenta much larger than those 
involved in typical nuclear ground state processes.  The study of nuclei 
is therefore directly related to the interpretation of high-energy 
experiments with nuclei from hadron colliders to fixed target experiments.

The understanding of nuclear effects is particularly relevant
for neutrino processes in which weak interaction with matter
requires the use of heavy nuclear targets in order to collect
significant number of interactions.
Therefore a reliable treatment of nuclear effects is 
important for interpretation of neutrino experiments and in some 
cases crucial for reducing systematic uncertainty.

In this paper we briefly review the results of recent studies of CL
nuclear DIS of Ref.\cite{KP04} and apply this approach to
calculate the (anti)neutrino differential cross sections of
charged-current (CC) interactions with nuclei.

\section{Nuclear structure functions}

The theoretical background of the analysis of Ref.\cite{KP04} involves
the treatment of a few different mechanisms of nuclear effects which
are characteristic for different kinematical regions.  Summarizing,
for the nuclear structure function of type $a=1,2,3$ we have
\begin{equation}
\label{FA}
F_a^A = F_a^{p/A} + F_a^{n/A} + F_a^{\pi/A} + \delta_{\rm coh} F_a^A,
\end{equation}
where $F_a^{p/A}$ ($F_a^{n/A}$) are the incoherent contributions
from the bound protons (neutrons) corrected for Fermi motion, nuclear 
binding and off-shell effects (impulse 
approximation). The term $F_a^{\pi/A}$ is a correction associated with 
scattering off the nuclear pion (meson) field.
The term  $\delta_{\rm coh} F_a^A$ is a correction due to coherent 
interaction of intermediate virtual boson with nuclear target.

The terms $F_a^{p/A}$ and $F_a^{n/A}$ are calculated as the proton and the 
neutron structure function averaged with the nuclear spectral function and 
corresponding kinematical factors. In particular, for $F_2^{p/A}$ we have%
\footnote{Eq.(\ref{IA}) is written for the kinematics of the Bjorken limit 
assuming that momentum transfer is along $z$ axis 
$q=(q_0,0_\perp,-|\bm{q}|)$. For more general expressions valid at finite 
$Q$ and further detail see Ref.\cite{KP04}.} 
\begin{equation}
F_2^{p/A}(x,Q^2)
= 
\left\langle \big(1+{k_z}/{M}\big)
F_2^p(x',Q^2,k^2)
\right\rangle,
\label{IA}
\end{equation}
where $k=(M+\ceps,\bm{k})$ and $x'=x/(1+(\ceps+k_z)/M)$ are the 
four-momentum of bound proton and its Bjorken variable. The averaging is 
taken over the proton spectral function which describes the distribution 
of bound nucleons over momentum and energy. Similar expression holds for 
the neutron term.

Nuclear Fermi motion and binding corrections provide a correct trend
of observed behavior of the ratios $\mathcal{R}_2=F_2^A/F_2^D$
(so-called EMC ratio) however the quantitative description of data is
missing in impulse approximation.  In a quantitative treatment it is
important to go beyond this approximation and take into account that
the structure functions of bound nucleons can be different from those
of the free nucleons. In Ref.\cite{KP04} this effect is modelled by
off-shell corrections, i.e. the dependence of structure functions on
$k^2$ in \eq{IA}.
Since characteristic energies and momenta of bound nucleons are small 
compared to the nucleon mass we treat off-shell effects as a linear 
correction in $k^2-M^2$ to the structure function of the on-shell nucleon
\begin{equation}
F_2(x,Q^2,k^2)=F_2(x,Q^2)\left(1+\delta f_2 \tfrac{k^2-M^2}{M^2}\right).
\label{os}
\end{equation}
The function $\delta f_2(x,Q^2)$ describes the relative off-shell effect. 
In analysis of Ref.\cite{KP04} this function is treated phenomenologically.
In particular, we assume 
$\delta f_2$ to be independent of $Q^2$ and parametrize its $x$ 
dependence as
\begin{equation}
\delta f_2=C_N(x-x_1)(x-x_0)(h-x),
\label{del:f}
\end{equation}
where $0<x_1<x_0<1$ and $h>1$.

It should be noted that because of binding the nucleons do not
carry all of the light-cone momentum of the nucleus and the momentum
balance equation is violated in the impulse approximation thus indicating 
the presence of non-nucleon degrees of freedom in the problem. In 
Ref.\cite{KP04} we consider a correction to nuclear structure function due 
to scattering off meson fields in nuclei in the convolution approximation
\begin{equation}\label{SF:pi}
F_2^{\pi/A}=\int_x \ud y f_{\pi/A}(y) F_2^{\pi}(x/y,Q^2),
\end{equation}
where $f_{\pi/A}(y)$ is the distribution of nuclear pion excess in a 
nucleus and $F_2^{\pi}$ is the pion structure function. The distribution 
function $f_{\pi/A}(y)$ is calculated using the constraints from equations 
of motion for interacting pion-nucleon system. By using the light-cone 
momentum balance equation we effectively constrain the contribution from 
all mesonic fields responsible for nuclear binding.

In the small-$x$ region the coherent effects in DIS are relevant
(for a review see Ref.\cite{Piller:1999wx}). These 
effects are associated with the fluctuations of intermediate virtual boson 
into quark-gluon (or hadronic) states. At small $x$ an average time of 
life of such fluctuation is significantly larger than the average distance 
between bound nucleons. For this reason the virtual hadronic states 
undergo multiple nuclear interactions while traversing a nucleus that 
causes nuclear shadowing effect. The rate of this effect depends on the 
scattering amplitude of the virtual hadronic states off the nucleon. In 
our approach we model the interaction of virtual hadronic states with the 
nucleon as scattering of a single state with effective cross section 
$\bar\sigma$. In this approximation the relative nuclear correction to the 
structure function 
is determined by the corresponding 
correction to effective cross section $\bar\sigma$. 
In our studies \cite{KP04} the effective cross section 
$\bar\sigma$ for the scattering off the nucleon 
is treated phenomenologically and parametrized as
\begin{equation}
\bar\sigma = \sigma_1 + \frac{\sigma_0 - \sigma_1}{1+Q^2/Q_0^2}
\label{sig:eff}
\end{equation}
where the parameters $\sigma_0$ and $\sigma_1$ describe low-$Q$ and
high-$Q$ limits while the scale $Q_0$ controls transition region.  The
nuclear corrections to effective cross section
were calculated using the Glauber--Gribov multiple
scattering theory.


We used the outlined approach in the analysis of data
on the ratios $\mathcal{R}_2(A/B)=F_2^A/F_2^B$ of structure functions
of two nuclei. The general goal was to develop a quantitative model of
nuclear structure functions which, from one side, would include the major
mechanisms of nuclear scattering and, from the other side, would
describe the existing data with acceptable accuracy.  We analyze data
on $\mathcal{R}_2$ for a variaty of targets from D to Pb for a wide
kinematical region (for more detail see Table~1 of Ref.\cite{KP04}).
%
%
From preliminary analysis of data we observed strong correlations between 
some of the model parameters.
In order to reduce the number of free parameters we used 
additional constraints. In particular, the parameters $h$ and $x_0$ turned 
out to be fully correlated and related as $h=1+x_0$. The parameter $x_1$ is
strongly correlated with $C_N$ and $Q_0$. We performed several fits with 
different fixed values of $x_1$ and found that $x_1=0.05$ corresponds to 
the lowest $\chi^2$ and provides a good cancellation between shadowing and 
off-shell correction to the normalization of nuclear valence quark 
distribution.
From preliminary fits the best fit value of $\sigma_1$ was consistent
with zero and we fixed $\sigma_1=0$ in the final fit. The parameter
$\sigma_0$ was fixed to 27\,mb (averaged meson-nucleon total cross
section in the vector meson dominance model
\cite{Bauer:iq,Piller:1999wx}) in order to reproduce the
photoproduction limit. The remaining parameters $C_N$, $x_0$ and $Q_0$
were adjusted to reproduce data. The global fit of Ref.\cite{KP04} to
all data results in $C_N = 8.1\pm 0.3\pm 0.5$, $x_0 = 0.448\pm
0.005\pm 0.007$, $Q_0^2 = 1.43\pm 0.06\pm 0.2\gevsq$ with $\chi^2{\rm
/d.o.f.}=459/556$ (the last error is the estimate of systematic or
theoretical uncertainty). In order to test the model we performed a
number of fits to different sub-sets of nuclei in the region from
$^4$He to ${}^{208}$Pb. The results are compaitible within the
uncertainties with the result of the global fit thus indicating an
excellent consistency between the model and the data for all nuclei.

\section{Neutrino-nucleus inelastic cross sections
}
\label{sec:nu}

In this paper we apply the developed model of nuclear structure
functions to calculate the differential cross
sections of (anti)neutrino inelastic scattering. 
In contrast to the CL scattering, which is driven by
electromagnetic interaction, neutrino interactions are characterized
by the presence of both the vector current (VC) and the axial current
(AC) contributions.  The VC--AC interference gives rise to P-odd and
C-odd terms in the cross section which are described by the structure
function $F_3$, which is absent in CL scattering. 
In contrast to VC the AC is not conserved. For this
reason the AC contribution plays important role and even dominates the
(anti)neutrino cross sections at low $Q$. In this region the terms due
to non conservation of the AC are most important for the longitudinal
structure function $F_L$ which is determined by the divergence of the
current (Adler theorem \cite{Adler:1964yx}). 
At low momentum transfer the divergence of the AC is
dominated by the pion field (PCAC) that allows us to calculate the
leading term at low $Q^2$:
$F_L^{\textsc{pcac}}=f_\pi\sigma_\pi(s,Q^2)/\pi$, where $f_\pi=0.93\, 
m_\pi$ is the pion decay constant and $\sigma_\pi$ is the total 
pion-nucleon (nucleus) cross section with the center-of-mass energy 
squared $s=M^2+Q^2(1/x-1)$.
It should be also noted that $\sigma_\pi$ describes interaction of an 
off-shell pion with momentum $q$ and the virtuality $-Q^2$.
Although the contribution from the VC to $F_L$ vanishes at low $Q^2$,
it rises with $Q^2$ and has to be taken 
into account at $Q^2\sim 1\gevsq$ as well as for higher values of momentum 
transfer.
We explicitly separate the PCAC term and write the full $F_L$ as
\begin{equation}\label{FL}
F_L(x,Q^2)=F_L^{\textsc{pcac}} f_{\textsc{pcac}}(Q^2) + 
\widetilde{F}_L(x,Q^2),
\end{equation}
where $\widetilde{F}_L$ incorporates the contributions from VC and 
non-PCAC terms from AC which vanish at $Q^2\to0$. In order to interpolate
between low and high $Q^2$ we introduced a form factor
$f_{\textsc{pcac}}=(1+Q^2/M_{\textsc{pcac}})^{-2}$, where
$M_{\textsc{pcac}}$ represents the scale controlling the PCAC
contribution.
Since the pion pole does not contribute to the structure
functions (because of transversity of the lepton current, see
Ref.\cite{Kopeliovich:2004px} and references therein) the scale
$M_{\textsc{pcac}}$ is not determined by the pion mass but rather
related to higher mass states like $a_1$ meson and $\rho\pi$ continuum.%
\footnote{In numerical calculations described below we used
$M_{\textsc{pcac}}=m_{a_1}$.  In Ref.\cite{Kopeliovich:2004px} it was
argued that the relevant scale is given by the $\rho\pi$ cut rather
than the $a_1$ pole. However, both values are close numerically.}
The term $\widetilde{F}_L$ dominates at high $Q^2$ at which it is
identified with the pQCD structure function with target mass (TMC) and
higher twist (HT) corrections.  In numerical applications we use the
PDFs and HT of Ref.\cite{a02} which were derived from the
analyses of CL DIS data. In order to evaluate $\widetilde{F}_L$ at low
values of $Q^2$ we apply polynomial extrapolation between pQCD
calculation at high $Q^2$ and the limit of $Q\to0$ with the matching
point at $Q^2=1\gevsq$ \cite{AKP06}.

\begin{figure}[htb]
\vspace{-4ex}\includegraphics[width=1.05\linewidth]{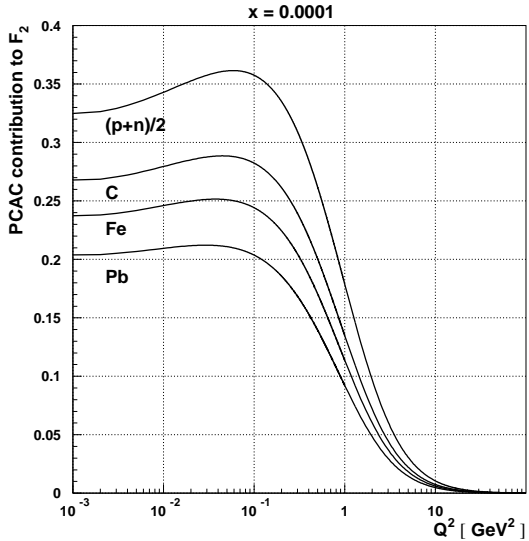}
\caption{%
The PCAC term in neutrino $F_2$ calculated as a function of
$Q^2$ for the fixed $x=0.0001$ \cite{KP06}. The results are shown for
the isoscalar nucleon (average over the proton and the neutron) and
for $^{12}$C, $^{56}$Fe and $^{208}$Pb nuclei.  }
\label{fig:FPCAC}
\end{figure}

The magnitude and the $Q^2$ dependence of the PCAC term in $F_2$ is 
illustrated in Fig.~\ref{fig:FPCAC}.
Note that the magnitude of the PCAC term decreases for heavy nuclei because of 
the nuclear shadowing effect for the pion cross section.

\begin{figure}[htb]
\includegraphics[width=1.05\linewidth]{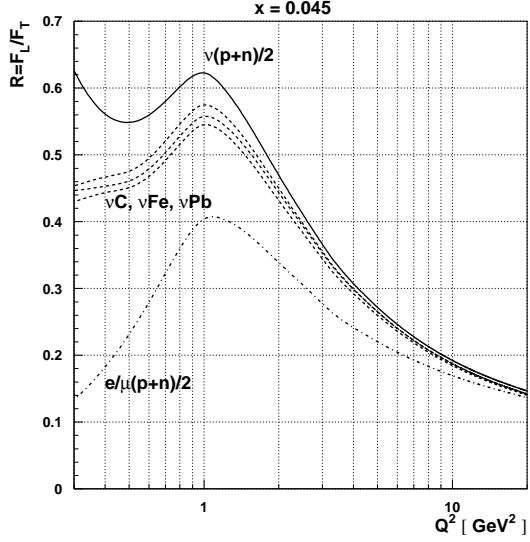}
\includegraphics[width=1.05\linewidth]{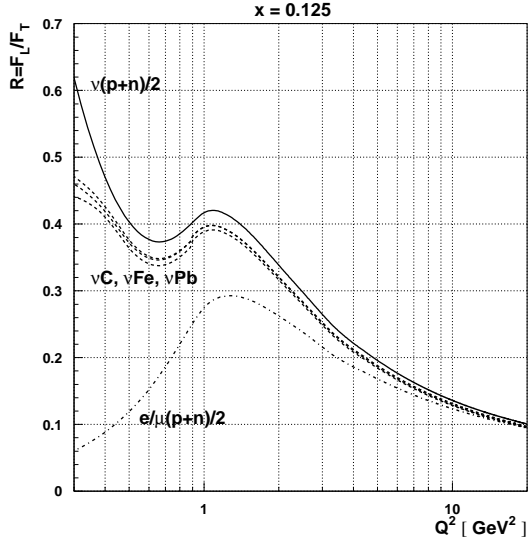}
\caption{Comparison of the ratio $R=F_L/F_T$ calculated for charged
leptons (dashed-dotted) and neutrino (solid) for the isoscalar nucleon
as a function of $Q^2$ at $x=0.045$ (upper panel) and $x=0.125$ (lower
panel). Also shown are the results of calculation of $R$ for neutrino
scattering off $^{12}$C, $^{56}$Fe and $^{208}$Pb (from top to bottom).
}
\label{fig:R}
\end{figure}

It is also important to note that for neutrino 
scattering the PCAC term leads to the rising ratio $R=F_L/F_T$ at low $Q^2$ 
in contrast to the CL case. This effect is illustrated in Fig.~\ref{fig:R} 
where $R(Q^2)$ was calculated for two fixed values of $x$ for the 
isoscalar nucleon (average over the proton and the neutron) and a number 
of nuclei \cite{AKP06,KP06}.

\begin{figure}[htb]
\includegraphics[width=1.05\linewidth]{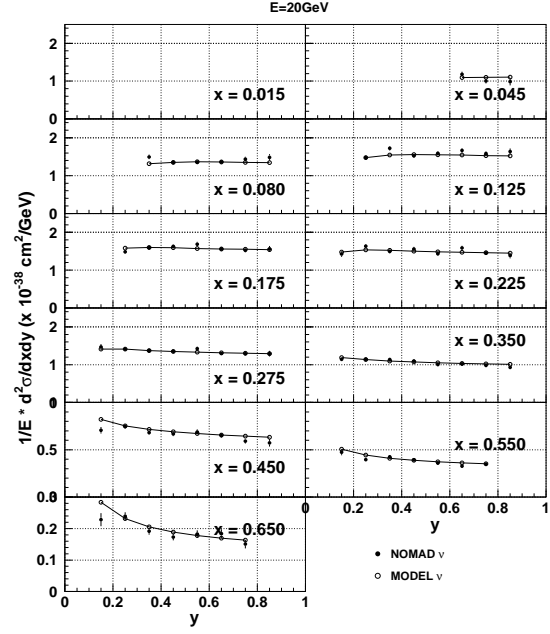}
\includegraphics[width=1.05\linewidth]{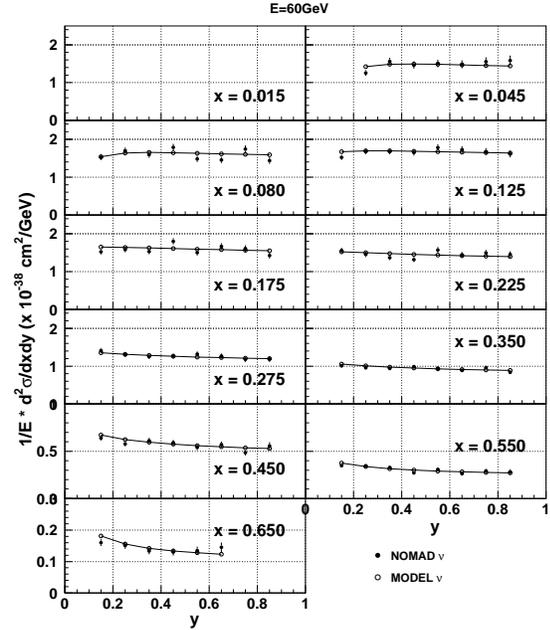}
\caption{Comparison of our calculations (open symbols) with 
NOMAD data \cite{RP-nuint05}
(full symbols) for neutrino differential cross-sections on ${}^{12}$C at 
$E=20\gev$ (upper panel) and $E=60\gev$ (lower panel).
}
\label{fig:NOMAD}
\end{figure}

\begin{figure}[htb]
\includegraphics[width=1.05\linewidth]{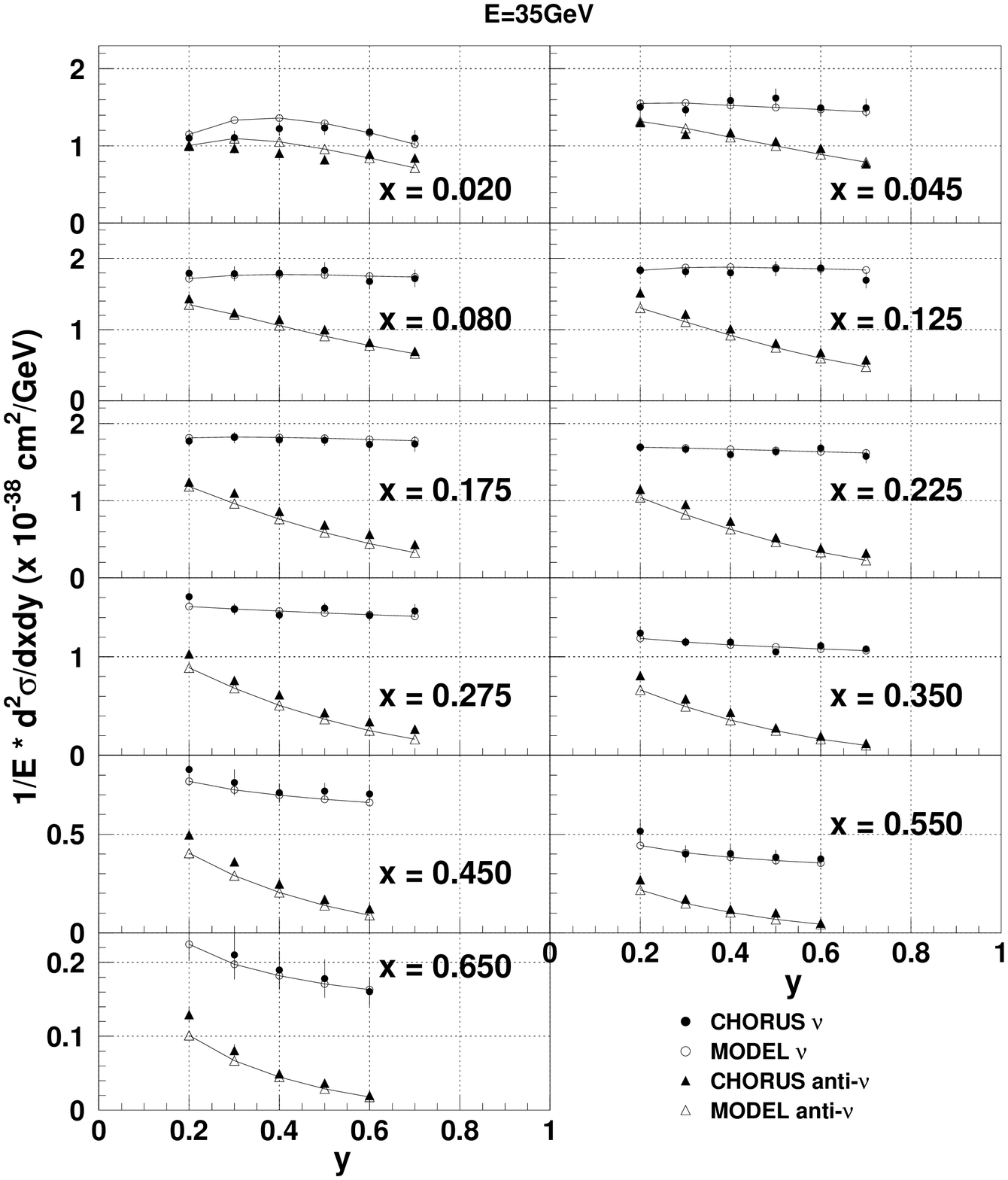}
\includegraphics[width=1.05\linewidth]{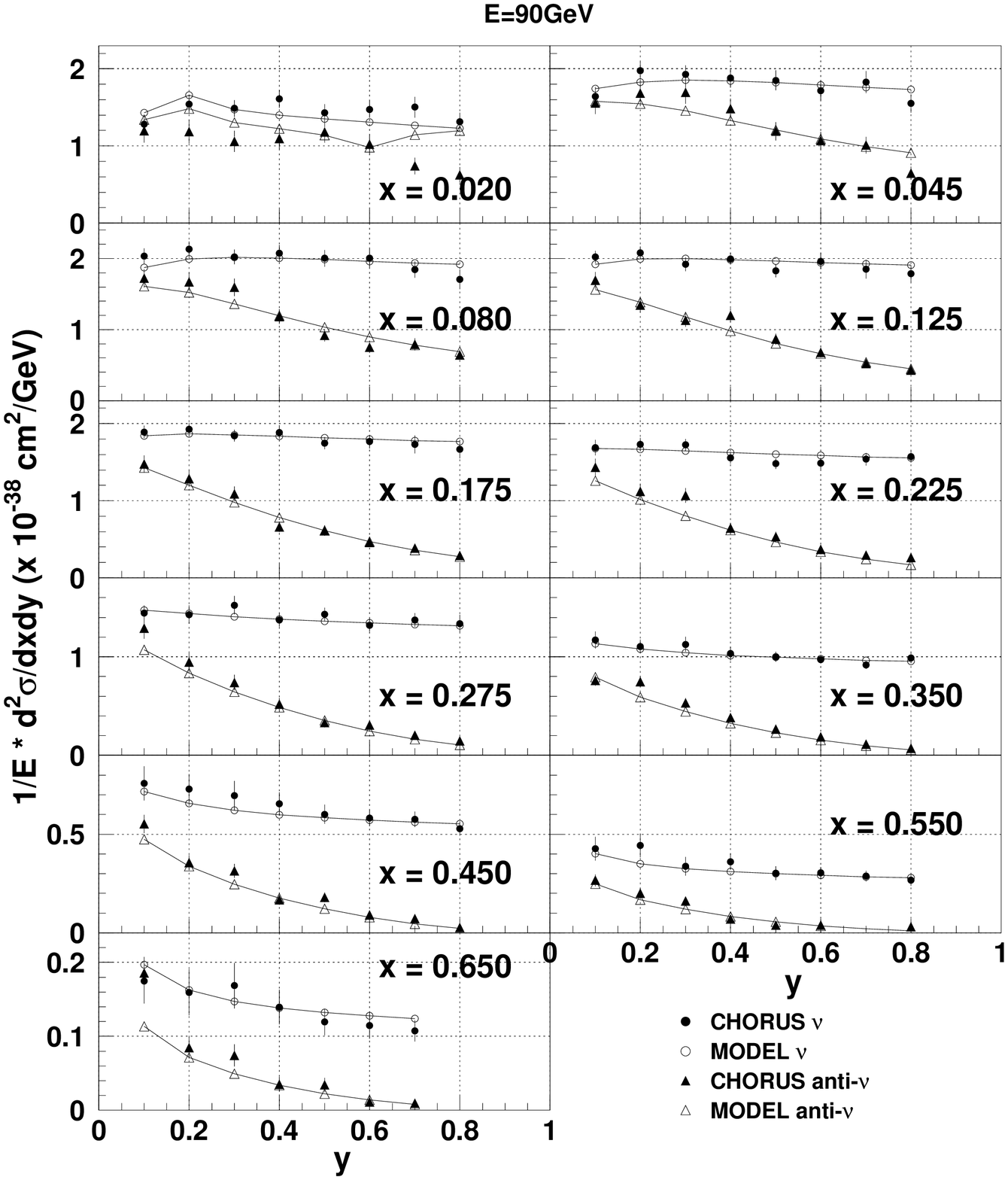}
\caption{Comparison of our calculation (open symbols) with 
CHORUS data \cite{Onengut:2005kv} 
(full symbols) for neutrino (circles) and antineutrino 
(triangles) differential cross-sections on ${}^{207}$Pb at 
$E=35\gev$ (upper plot) and $E=90\gev$ (lower plot).
}
\label{fig:CHORUS}
\end{figure}

\begin{figure}[htb]
\includegraphics[width=1.05\linewidth]{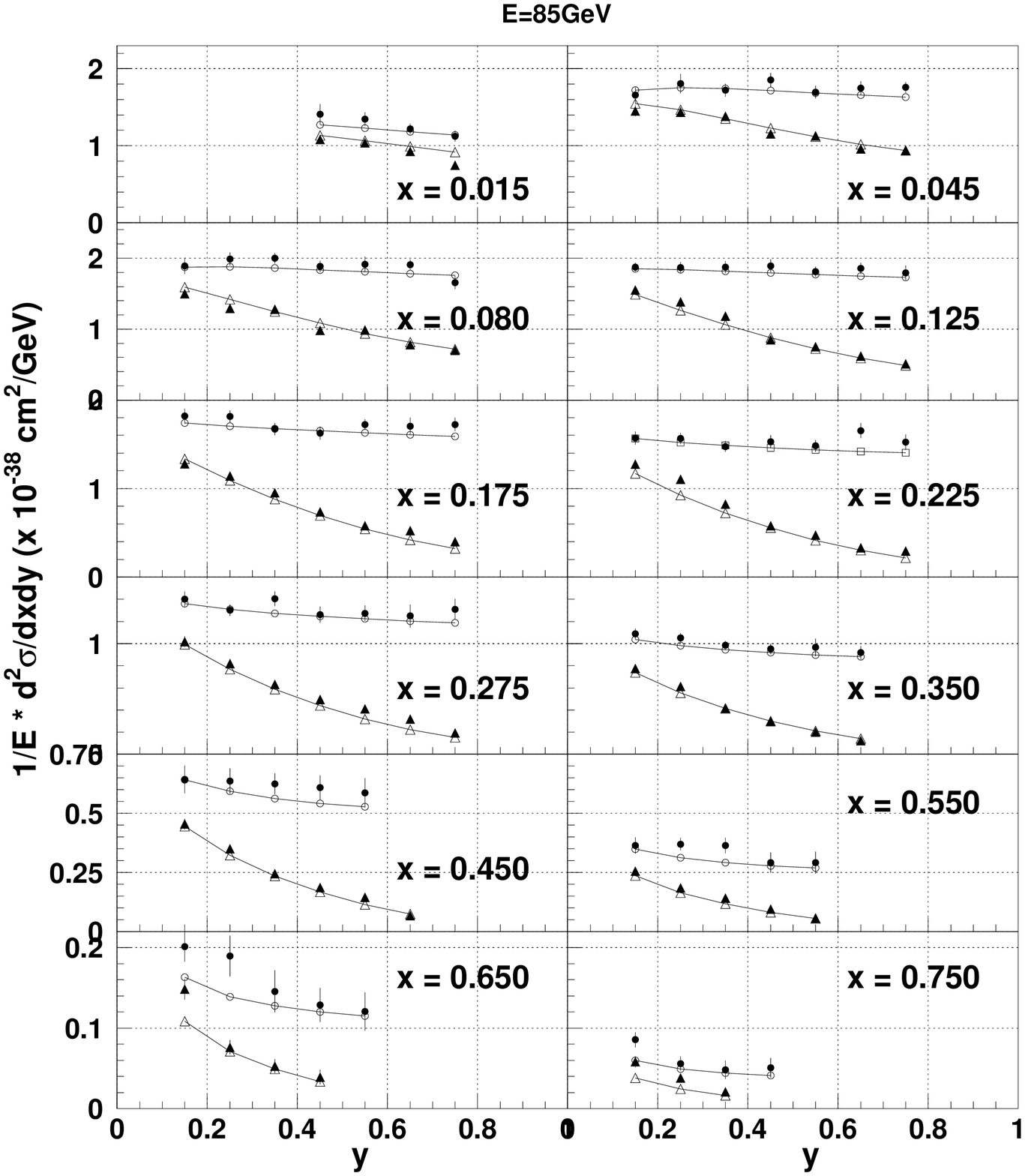}
\includegraphics[width=1.05\linewidth]{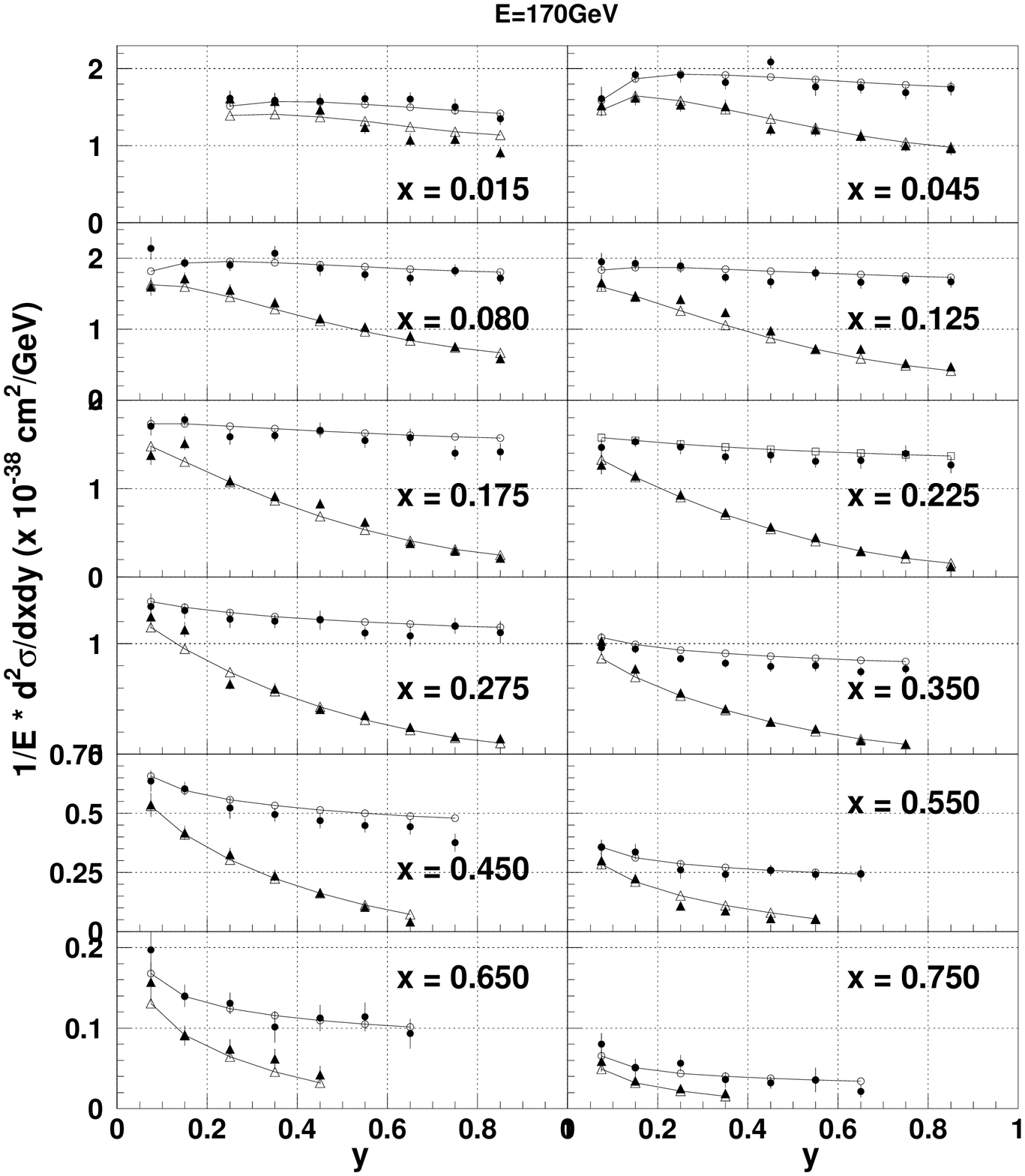}
\caption{Comparison of our calculation (open symbols) with NuTeV data 
\cite{nutev-xsec} (full symbols) for neutrino (circles) and antineutrino 
(triangles) differential cross-sections on ${}^{56}$Fe at 
$E=85\gev$ (upper plot) and $E=170\gev$ (lower plot).
}
\label{fig:NuTeV}
\end{figure}

In Figures~\ref{fig:NOMAD},~\ref{fig:CHORUS} and~\ref{fig:NuTeV} we
report the results of our calculations of (anti)neutrino differential
cross sections in comparison with the measurements for several values
of the (anti)neutrino energy from 20\:GeV to 170\:GeV \cite{KP06}.
The calculation of (anti)neutrino nuclear structure
functions uses the same NNLO PDFs and HT terms obtained from the
analyses of charged-lepton DIS data \cite{a02} and which were applied
in the analysis described above. 
The treatment of nuclear effects is based on the results of Ref.\cite{KP04}.
In the calculations we used the off-shell function
$\delta f_2$, derived from CL data in Ref.\cite{KP04},
for both $F_2$ and $F_3$ for neutrino and
antineutrino. 
Nuclear effects in the PCAC term are
controlled by the multiple scattering corrections to the pion cross
section $\sigma_\pi$. We also remark that nuclear pion excess
correction (\eq{SF:pi}) vanishes for $F_3$.
As a result, we observe a good agreement between the data and our
calculations for all examined nuclei that provides a good test of the 
model of Ref.\cite{KP04}. It should be noted that the data points at low $x$ 
bins, which typically have low $Q^2$ (for example for CHORUS data $Q^2\sim 
0.25 \gevsq$), are also reproduced by calculation. In this region the 
cross sections are dominated by the PCAC term. Thus our analysis supports 
the presence of significant PCAC effects in neutrino inelastic scattering 
at $Q^2\gg m_\pi^2$.

\section*{Acknowledgments}

S.K. is grateful to the organizers of the NuInt05
for support and hospitality.

\end{document}